\documentclass[reprint,prl]{revtex4-1}
\usepackage{amsmath,amssymb,braket}
\usepackage{graphicx}

\usepackage{xcolor}

\begin{document}
	
	\title{The Capacity of Quantum Neural Networks}
	\author{Logan G. Wright$^{1,2,}$} 
	\email{lgw32@cornell.edu}
	\author{Peter L. McMahon$^{1,2,}$}
	\email{pmcmahon@cornell.edu}
	\affiliation{1. School of Applied and Engineering Physics, Cornell University, Ithaca, NY 14853, USA}
	\affiliation{2. E.\,L. Ginzton Laboratory, Stanford University, Stanford, CA 94305, USA}

	\begin{abstract}A key open question in quantum computation is what advantages quantum neural networks (QNNs) may have over classical neural networks (NNs), and in what situations these advantages may transpire. Here we address this question by studying the memory capacity $C$ of QNNs, which is a metric of the expressive power of a QNN that we have adapted from classical NN theory. We present a capacity inequality showing that the capacity of a QNN is bounded by the information $W$ that can be trained into its parameters: $C \leq W$. One consequence of this bound is that QNNs that are parameterized classically do not show an advantage in capacity over classical NNs having an equal number of parameters. However, QNNs that are parametrized with quantum states could have exponentially larger capacities. We illustrate our theoretical results with numerical experiments by simulating a particular QNN based on a Gaussian Boson Sampler. We also study the influence of sampling due to wavefunction collapse during operation of the QNN, and provide an analytical expression connecting the capacity to the number of times the quantum system is measured.
	\end{abstract}
	\maketitle
	
	While they have been of interest since the 1990s \cite{Menneer1995,Kak1995,Behrman1996}, research on quantum artificial neural networks (QNNs) is expanding, invigorated by possible implementations on near-term hardware \cite{Farhi2018}. What is a QNN? Here we take a broad definition that encompasses emergent themes across a variety of proposals (and that, as with classical NNs, have a mostly historical connection to biological NNs). QNNs being considered for near term implementations are a class of variational quantum algorithms \cite{Farhi2014,Peruzzo2014,McClean2016,Benedetti2019}: quantum circuits having gates whose parameters are adjusted to perform a desired transformation on either classical or quantum information that is fed forward through the circuit. QNN architectures include ``quantized" classical neural networks  \cite{Schuld,Wan2017,Farhi2018,Rebentrost2018,Killoran2018,Steinbrecher2018,Beer2019,Tacchino2019,Cong2018,Carolan2019} and Boltzmann machines \cite{Wiebe2014b,Benedetti2016,Verdon2017,Adachi2015,Amin2018}, as well as (within our broad definition) schemes using quantum circuits with more opportunistic configurations that exploit trainable input-output mapping of near-term quantum circuits \cite{Mitarai2018,Wilson2018, Schuld2018,Schuld2019,Havlicek2019}. 
	
	What are QNNs good for? While there is not yet a consensus, some proposals have indicated potential quantum speed-ups or advantages in the size or training of the network (e.g., \cite{Ventura1998,Wiebe2014,Wiebe2014b,Arunachalam2017,Verdon2017,Verdon2018,Rebentrost2018}). Many are intended as general quantum learning machines capable of a broad range of tasks (e.g., \cite{Biamonte2008,Wan2017,Farhi2018,Killoran2018,Steinbrecher2018,Beer2019,Biamonte2019}). This possibility of generality is stimulating: classical machine learning has found widespread use in large part because it permits even inexpert users to experimentally discover algorithms for their diverse and specific tasks. Last, many QNNs appear suitable for implementation on noisy intermediate-scale quantum (NISQ) computing devices \cite{Preskill2018}, for which there is an ongoing search for suitable algorithms.
	
	Despite its promise, QNN development faces several challenges. While many QNN proposals have been made, comparing them remains challenging, especially for concrete hardware implementations. This is an important challenge to resolve due to the breadth of possible hardware implementations, spanning numerous platforms and both discrete and continuous-variable quantum systems. Training QNNs has emerged as another challenge \cite{Verdon2017,Verdon2018,Nakanishi2019,Bergholm2018,Schuld2019b,McClean2018}. Many QNN proposals for supervised learning rely on classical computers to perform the training that determines (classical-valued) parameters of the quantum circuit. The probabilistic nature of the quantum variables used imposes a trade-off between measurement time and training accuracy, and the optimization problem landscape poses challenges for gradient descent \cite{McClean2018}. Quantum Boltzmann machines \cite{Amin2018}, especially quantum training of Boltzmann machines \cite{Wiebe2014b,Verdon2017} are promising, but many other proposals \cite{Mitarai2018,McClean2018,Beer2019,Nakanishi2019,Bergholm2018,Schuld2019b,Carolan2019,Mitarai2018,Nakanishi2019,Bergholm2018,Schuld2019b,Parrish2019}, may help to address these challenges. Finally, extrapolation of small-scale QNNs is challenging: typically researchers cannot confidently predict how or if scaled-up QNNs will be useful. 
	
	Our main result is inspired by information-theoretic models of classical NNs \cite{Vapnik2015,Mackay2005, Tishby2015,Shwartz-Ziv2017,Friedland2017,Baldi2018,Friedland2018,Baldi2019}, and is a generalization of the memory capacity  \cite{Cover1965,Gardner1987,Kinzel1998,Mackay2005,Friedland2017,Baldi2018,Friedland2018,Baldi2019} that is closely related to the Vapnik-Chervonenkis dimension \cite{Vapnik2015,Shalev-Shwartz2013,Vapnik2000}. These measures aim to quantify the complexity of representation a given classifier can achieve. The capacity is a simplistic measure; it is useful in the following senses \cite{Vapnik2000,Shalev-Shwartz2013}. First, with a suitable learning method a classifier may learn a range of tasks whose maximum complexity is bounded by its capacity. Second, the capacity describes the information that must be provided to the classifier in training to ensure generalization, i.e., to avoid overfitting. Preventing over-fitting can also be accomplished by techniques that automatically restrict capacity, such as parameter regularization. 
	
	Adapting the notion of capacity to QNNs allows us to quantify the range of tasks a QNN may be used for, and the training requirements to ensure good generalization.  Quantum learning machines intuitively support much more complex models than classical ones, so managing overfitting may ultimately be an important challenge. On the other hand, learning \textit{arbitrary} quantum tasks requires enormous capacity: arbitrary quantum models may be exponentially more complex than those required for classical tasks.  
	
	Here, by considering the memory capacity in information units (bits or qubits), we show that a simple capacity measure can be universally defined to encompass all possible learning machines: classical, quantum and hybrids. To relate this capacity to the learning machine's physical implementation, we introduce a capacity inequality that bounds capacity by the information that can be imparted by training to the machine's trainable parameters. We anticipate this approach will be useful for establishing rough guidelines for QNN design and application, similar to how capacity has been used to understand classical NNs \cite{Cover1965,Gardner1987,Kinzel1998,Mackay2005,Friedland2017,Baldi2018,Friedland2018,Baldi2019}. 
	
	The idea of memory capacity as a limit of learnable complexity can be intuitively understood as follows. Consider a complex task such as a sequence of errands. The instructions for completing such a task may be partially redundant, but there is a minimum amount one must remember in order to execute the task flawlessly. In the limit of a completely random sequence of errands, the impossibility of universal lossless compression implies that a learner must remember all the information provided by the instructions, $T$, exactly. Let us call the size of the largest random task a learner can ``learn" (memorize) its capacity, $C$. A necessary condition if there are $T$ bits to memorize is that the learner possess the ability to meaningfully change at least $T$ bits of its degrees of freedom as it learns. This follows directly from the pigeonhole principle. When trained for a general task specified by data $T>C$, a learner cannot memorize all $T$ bits. Instead, to minimize training error it must learn a compressed model by exploiting patterns within the data. 
	
	\textit{Definitions:} More formally, we define the memory capacity, $C$, of a learning machine as the largest $Nm$ bits of $N$, $m$-bit labels for which it can learn to label $N$ general position inputs (which may be quantum and/or classical, depending on the machine) by any combination of $N$, $m$-bit labellings. General position implies uniformly-distributed distinct random points, drawn from the across the range of acceptable inputs to the learning machine \cite{Mackay2005}. Essentially, $C$ is the largest $Nm$-bit RAM the learning machine can be trained to emulate. We also introduce $W$, defined as the amount of information which can be stored through training in the trainable parameters of the learning machine.  For example, consider a learning machine whose $N_{\text{w}}$ classical trainable parameters $w_i$ can each be trained with certainty to take on one of $M_i$ distinct levels. Then $W = \sum_{i=1}^{N_{\text{w}}}\log_2M_i = \sum_{i=1}^{N_{\text{w}}}b_i$ bits. 
	
	The rationale for these definitions is as follows. Consider the task of exactly learning a random map, i.e., a map from $N$ uniformly-distributed, distinct random inputs to $N$ uniformly-distributed, distinct random $m$-bit labels. Because this data is incompressible, a learning machine must be able to learn a model whose complexity, measured in terms of the information needed to describe it, is no less than the amount of information contained in the random labels. This statement assumes, however, that the learning machine's initial state (prior to learning) is completely uncorrelated with the random map; it may be possible to find a particular task or subset of $T$-bit tasks that the learning machine is (by chance or design) initialized to perform, or can learn with only small parameter adjustments. Such a machine would not necessarily be able to learn arbitrary $T$-bit tasks, nor function as a $T$-bit RAM. Therefore, for $N$ inputs in general position, all possible labellings of $N$, $m$-bit numbers should be learnable exactly in order to ensure $C\ge Nm$ bits. 
	
	Our main result is to relate $C$ for an arbitrary learning machine to its physical implementation including its learning algorithm, via $W$, through the inequality: 
	\begin{equation} \label{equation1}
	C \le W.
	\end{equation} 
	
	The proof is simple and has been essentially stated earlier in this paper and by others in more specific contexts, but for completeness we give it in full here. We will for now consider devices producing classical outputs, the quantum case will be considered subsequently.
	
	\textit{Proof:} If a learning machine is able to learn a model that requires $T$ bits to specify, then the machine must possess degrees of freedom that can store at least $T$ bits of information, those degrees of freedom must be trainable, and its training must be able to store at least $T$ bits of information in them. The first requirement is just due to the pigeonhole principle, and the second and third requirements are necessary to describe parameterized learning machines. Thus $C$, the memory capacity or maximum model complexity of the learning machine, can be no larger than the information that can be stored in its trainable parameters through learning, $W$, i.e., $C \le W$.
	
	\begin{figure}
		\centering
		\includegraphics[width=\columnwidth]{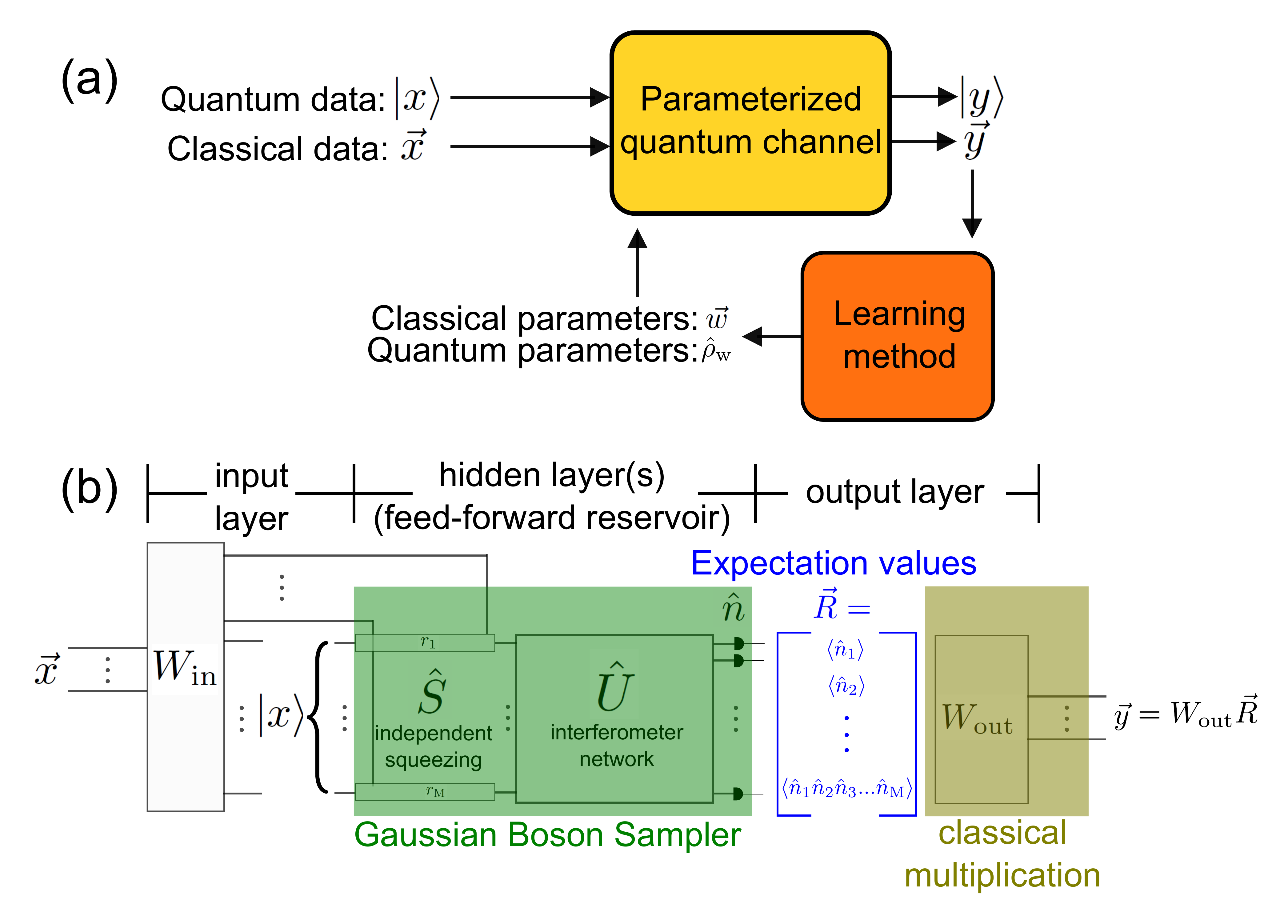}
		\caption{(a) Schematic of a general feed-forward QNN, a parameterized quantum channel (which could include unitary and/or dissipative quantum evolutions, classical data processing, ancillary parameter states, etc.) which is trained in a supervised fashion to optimize the classical and quantum parameters $\vec{w}$ and/or $\hat{\rho}_{\text{w}}$ so that the QNN best approximates the transformation implied by the training data. (b) Schematic of a feed-forward quantum reservoir computer based on a Gaussian Boson Sampler. For classical tasks considered here, $\ket{x} = \ket{0}$ and data is then encoded through the squeezing parameters, and for all tasks we take $W_{\text{in}}$ to be the identify matrix.   }		
	\end{figure}
	
	Until now, our considered learning machine has been general. Equation \ref{equation1} \textit{is} general: it applies to any learning machine with trainable parameters. We now consider the parameterized quantum channel depicted in Fig. 1a, which describes a general feed-forward artificial QNN. The device maps inputs -- a tuple of quantum and classical data -- to outputs that may also contain quantum and classical parts, i.e., $(\ket{x},\vec{x}) \mapsto (\ket{y},\vec{y})$. Supervised training of the QNN uses input-output pairs as training data (e.g., the $x$ and $y=f(x)$ values from a nonlinear function) or quantum channel (e.g., a unitary quantum circuit or dissipative evolution), and attempts to optimize the QNN's parameters to make the QNN's outputs for each input match the training set. In addition to depending on the QNN architecture (the layout of the QNN and its trainable parameters), $C$ and $W$ also depend on the execution and training protocols (which include, e.g., the input data encoding and learning method). Although $C$ and $W$ depend on these details, Eqn. \ref{equation1} applies universally, regardless of whether the learning machine and/or training involves quantum, classical, or hybrid operations, whether the trained parameters are classical or quantum, how many uses of the QNN (or repeats of the input data) occur per input, or how the data is encoded. 
	
	The interpretation of Eqn. \ref{equation1} in the case of quantum parameterization and/or quantum tasks can be understood as follows. If the QNN's trainable parameters are quantum states that can be learned to some finite ($m$-bit) precision, they comprise a total density operator $\hat{\rho}_{\text{w}}$ in a Hilbert space of dimension $D$, which can be described by a matrix containing $D^2-1$ real $m$-bit numbers. That is, $W=m(D^2-1)$ bits. Using this same technique of considering quantum objects as $m$-bit approximations of their classical matrix representations, we can similarly interpret $C$ and quantum tasks. Quantum tasks, such as preparing states or learning a quantum circuit, are unitary approximation tasks. For example, if the optimal quantum model for a given quantum task is an $M$-qubit unitary, it may be described by a matrix with up to $2^{2M}-1$ real degrees of freedom. An $m$-bit approximation of the unitary would be one in which each matrix element is correct to within $m$-bit precision. To learn an $m$-bit approximation of an \textit{arbitrary} $M$-qubit unitary, one requires a QNN with $C \ge m(2^{2M}-1)$ bits. Let us emphasize that $C$ is not equivalent to computational hardness or ``power". For example, a circuit initialized to implement Shor's algorithm exactly can perform the hard task of factoring even with an arbitrarily small $C$. On the other hand, should we wish to train it to perform other tasks, we can only count on it learning tasks within some ``radius" of $C$ bits from the initial unitary (i.e., unitaries accessible by changing parameter values by $C$ bits). Finally, our definitions of $C$ and $W$ above apply to learning machines producing classical outputs. For machines producing quantum outputs, a natural (but cumbersome) generalization is to task the device with mapping each of the $N$ inputs to $N$ label states, on which tomography is performed to arbitrary precision (so that errors in the measured matrix elements are dominated by the imprecision of the quantum learning machine). The information content $m$ of each label state is then quantified as the summation of the machine-limited bit precision across the density-matrix elements of each label state, i.e., $\sum_{ij}\log_2{e_{ij}}$, where $e_{ij}$ is the error in the reproduction of the label state's $(i,j)$-th matrix element. 
	
	As a concrete physical example, we consider a simple QNN, a feed-forward reservoir computer (fQRC) based on a Gaussian Boson Sampler (GBS). The GBS-fQRC (Fig. 1b) consists of an input layer, where quantum or classical data is injected into a reservoir circuit (a GBS), and an output layer that follows the reservoir. The reservoir circuit is feed-forward and untrained; it serves to transform the input data into a feature space, similar to the scheme proposed in Ref. \cite{Schuld2019}, producing a feature vector $\vec{R}$ for each input. The final output result $\vec{y}$ follows from linear matrix multiplication to select a linear combination of features to approximate the desired function, $y=W_{\text{out}}\vec{R}$. Thus, this QNN always produces a classical output regardless of whether its inputs are quantum or classical data. Training consists of a linear regression to optimize the matrix $W_{\text{out}}$ over a training set (see SM for more details \cite{SM}). This is a quantum single-hidden-layer NN, analogous to an ``extreme learning machine" (ELM) \cite{Huang2004} in the classical literature. Recent works have explored quantum reservoir computers (QRC) \cite{Fujii2017,Negoro2018,Nakajima2018,Kutvonen2018,Ghosh2019,Chen2019} that adapt the concept of reservoir computing \cite{Jaeger2001,Maass2002} to quantum dynamical systems. The GBS-fQRC simplifies this approach by considering a feed-forward device, implemented with a GBS. Boson Samplers and GBS are based on the propagation of photons or squeezed states through a multimode interferometer network, followed by photon detection \cite{Aaronson2011,Lund2014,Hamilton2017,Huh2015}. Our primary motivation in proposing the GBS-fQRC here is that it is a simple QNN, amenable to both theoretical and experimental studies (however, GBS appears promising for applications as well \cite{Bradler2018,Arrazola2018,Schuld2019c}). 
	
	\begin{figure}
		\centering
		\includegraphics[width=\columnwidth]{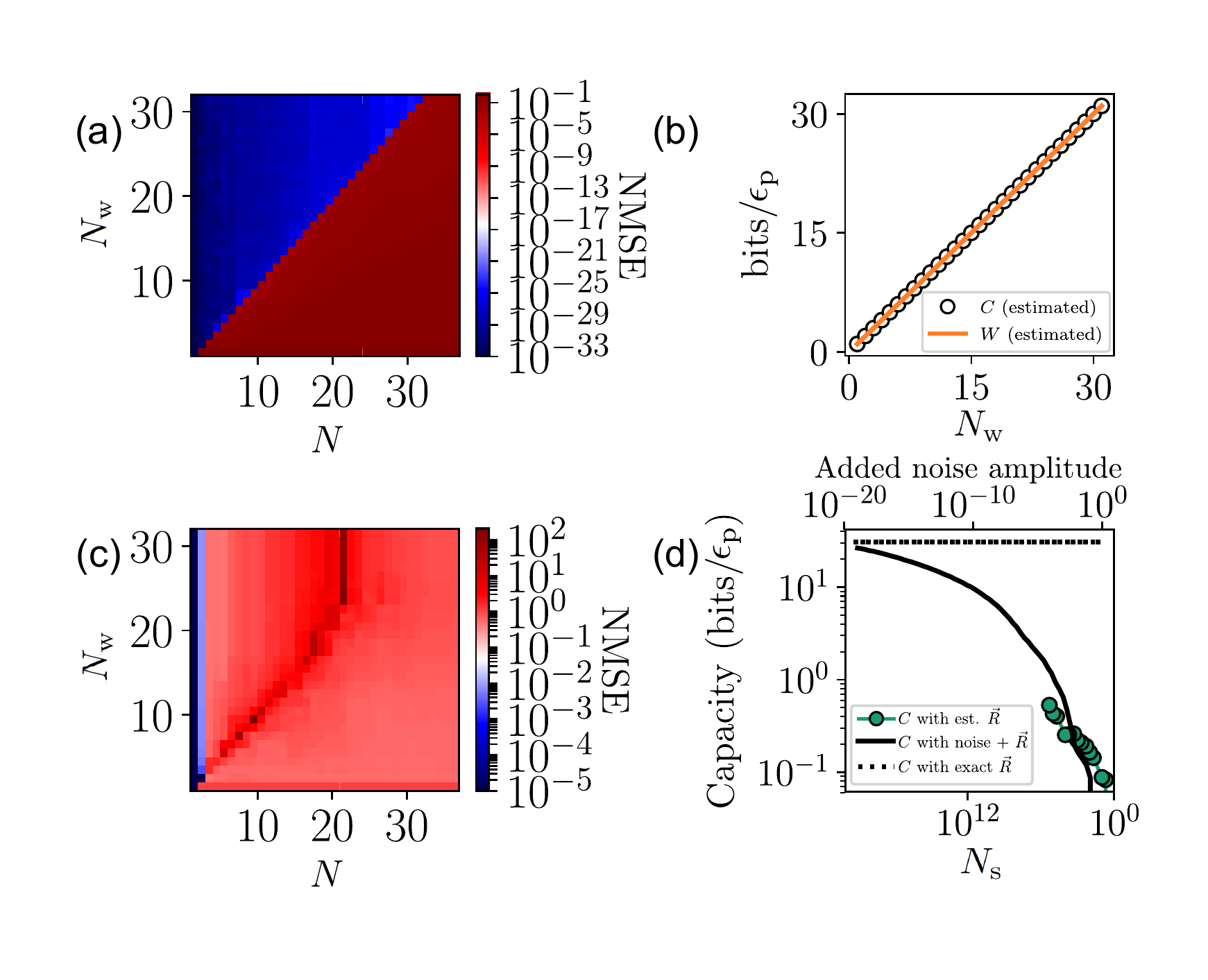}
		\caption{Estimation of the capacity of the GBS-fQRC. (a) shows the average normalized mean-squared-error (NMSE) for classification of $N$ double-precision numbers with a GBS-fQRC containing $N_{\text{w}}$ trainable parameters, where the simulation has been given access to the exact expectation value vector $\vec{R}$. (b) shows the estimated capacity and $W$ for this version of the QNN. The units are relative to the numerical-noise-limited bit depth $\epsilon_{\text{p}} \approx 45$ bits (for details, see Supplemental Material \cite{SM}). (c) shows the same plot as (a) when $N_{\text{s}}$=$10^5$ samples are used to estimate $\vec{R}$. (d) shows the estimated capacity versus the number of measurements used in the estimate, $N_{\text{s}}$ (bottom axis), and versus added noise amplitude (top axis).  
		}		
	\end{figure}
	
	\begin{figure}
		\centering
		\includegraphics[width=\columnwidth]{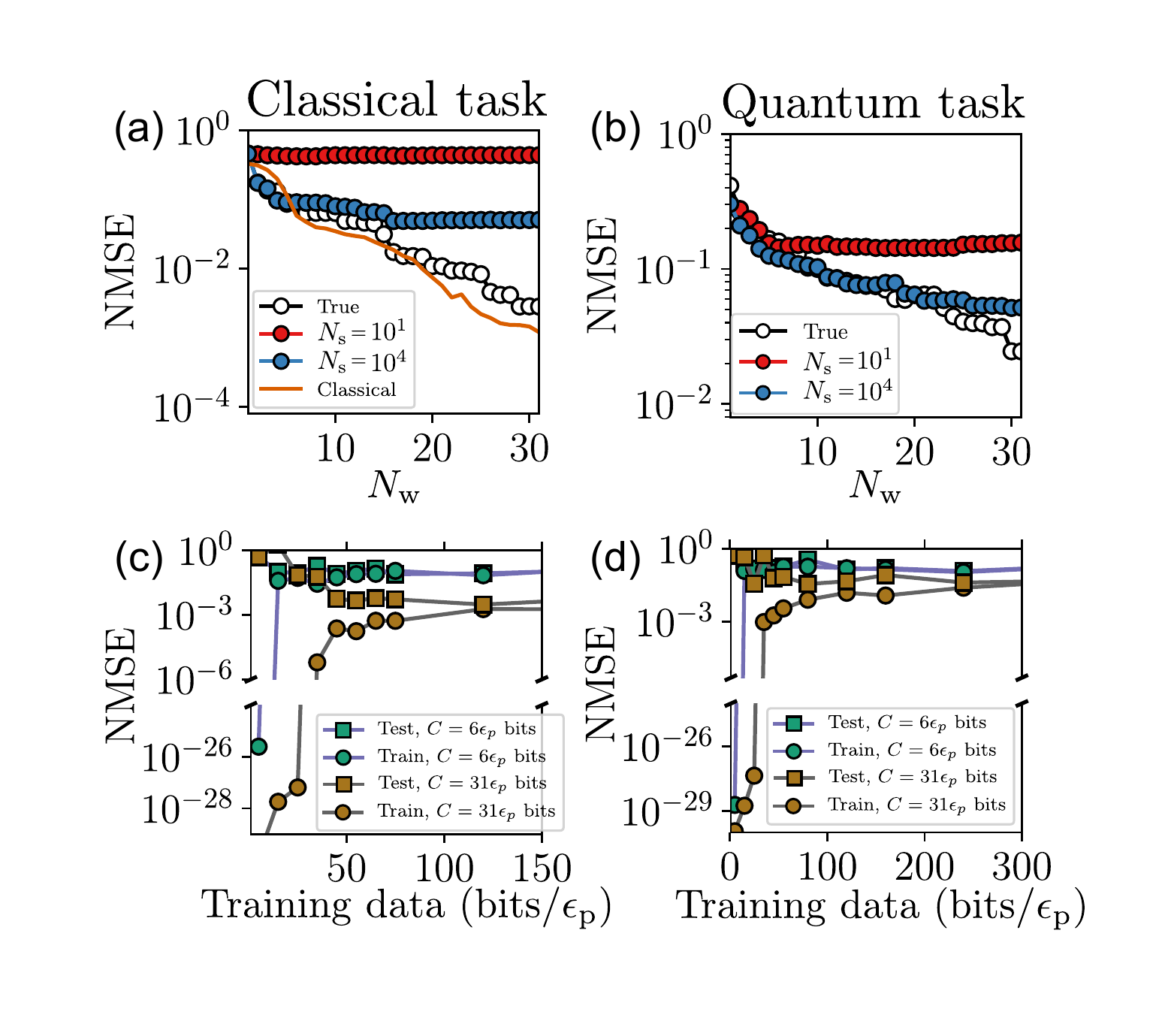}
		\caption{Performance and generalization of a QNN. (a) Performance of a 5-mode GBS-fQRC in computing a nonlinear function (quantified as normalized-mean-squared-error on a validation data set) for varying number of trained parameters, $N_{\text{w}}$. $N_{\text{s}}$ is the number of samples used to estimate the quantum variables in $\vec{R}$. The orange solid line shows the same for a classical ELM (for details see SM \cite{SM}). (b) Performance on a quantum task of predicting a complicated operator's expectation value on an injected quantum state, versus the number of trained parameters. (c) Generalization occurs when training data exceeds $C$, as can be seen by plotting training and test error versus the amount of data used to train for classical and (d) quantum task. Simulations were performed using the Strawberry Fields and QuTiP Python packages \cite{Killoran2018a,qutip}.}		
	\end{figure}
	
	To measure $C$ for a QNN, we need to determine the largest RAM it can emulate. That is, given $N$ distinct random inputs distributed uniformly across the relevant input space, the network is trained to classify the $N$ inputs into $N$ distinct, $m$-bit outputs. If perfect ($m$-bit precision-limited) classification is achieved for all possible $N$ $m$-bit labellings, then $C \ge Nm$. Thus, to find $C$, $Nm$ is increased until training error increases. A systematic measurement of $C$ is prohibitive even for small networks, but we can estimate $C$ by trying a small fraction of possible labellings. Fig. 2 shows the capacity estimate for a 5-mode GBS-fQRC. The estimated capacity linearly increases with the number of trainable parameters, increases with the number of samples used to make estimates of the expectation values, and decreases with the amount of artificial noise added to the true expectation values. It can be proven that the fQRC can theoretically achieve $C=W$, although we find that simulation precision prevents this for both classical and quantum devices \cite{SM}. 
	
	The GBS-fQRC can be trained to perform quantum and classical tasks, and in each case the performance depends on $C$. Fig. 3 shows the normalized mean-squared error (NMSE) on a test data set (i.e., a different set than was used to train, drawn from the same range of values) for a 5-mode GBS-fQRC trained to compute the function of input classical numbers ${x_i}$, $f({x_i}) :=(\sum_{i=1}^5x_i)^4+\sum_{i=1}^5x_i^3$, which is an arbitrarily chosen nonlinear function. In Fig. 3a, the NMSE is plotted versus the number of expectation values considered in the output layer (equal to the number of trainable parameters here, since the output $y$ is a scalar). As the number of trained parameters increases, the NMSE on the task drops. Similar behavior is observed for a variety of continuous functions \cite{SM}. If, instead of using the ``true" expectation values directly from quantum simulations, we estimate them by performing $N_{\text{s}}$ simulated measurements, we observe similar trends, except that the performance improvement saturates at a number of parameters that depends on $N_{\text{s}}$. This is a consequence of the reduced capacity due to reduced precision of training and execution. As a reference, the continuous curve in Fig. 3a shows the NMSE for the classical analogue, an ELM with a varying number of hidden nodes (which for this scalar-output task is equal to $N_{\text{w}}$).
	
	Similar findings are obtained for quantum tasks. As an example quantum task, we consider the prediction of the expectation value of an operator acting on injected quantum states. An injected state $\ket{x}$ is randomly selected state of the form $\ket{\psi_1} \otimes \ket{\psi_2} \otimes \ket{\psi_3} \otimes \ket{0} \otimes \ket{0}$ where each $\ket{\psi_i}$ is randomly selected to be a Fock, cat, coherent or thermal state, with a randomly-selected mean photon number between 0 and 1 (or equal to 1 or 2 for the Fock state). After passing the state through the GBS, the expectation-value vector $\vec{R}$ is measured, and multiplied by the trained output matrix to produce the expectation value prediction. Fig. 3b shows the NMSE for a network trained to compute the absolute value of the arbitrarily chosen operator $\hat{O} = \hat{x}^4_1 \hat{x}^2_2 \hat{x}^4_3 $, where $\hat{x}_1:=\hat{x}\otimes \hat{1} \otimes \hat{1} \otimes \hat{1} \otimes \hat{1}$ etc., versus $N_\textrm{w}$. Like the previous example, this task is chosen for concreteness; the GBS-fQRC can be trained to predict other expectation values or perform other quantum tasks, like state classification \cite{Ghosh2019,SM}. 
	
	Both Fig. 2 and 3 illustrate the impact of precision in the trainable parameters. Consider that $W=\sum_{i=1}^{N_\textrm{w}}\log_2(M_i)=\sum_{i=1}^{N_\textrm{w}}\log_2(\text{max}(w_i)/\delta w)$, where $\delta w$ and $\text{max}(w_i)$ are the precision and maximum possible value of the $i$th parameter, $w_i$. For fQRC, $y=W_{\textrm{out}}\vec{R}$. Thus, if the noise in $\vec{R}$ is $\propto \frac{1}{\sqrt{N_\textrm{s}}}$, then $W$ can be simplified to $W \propto N_\textrm{w}\log_2(N_\textrm{s})$. Assuming the upper bound of $C=W$ and neglecting all other precision-limiting effects, this implies that $C$ increases linearly with $N_\textrm{w}$, but only logarithmically with the number of samples $N_\textrm{s}$ used to estimate $\vec{R}$.
	
	We are interested in learning models that generalize well, a quality that is often at odds with complexity. For the tasks we consider here, generalization occurs when the amount of training data exceeds $C$, ensuring that memorization is no longer a good strategy to minimize training error. Figs 3c-d show the training and test error versus the uncompressed information provided in training, $T=N_{\text{T}}b_{\text{T}}$ bits (where $N_{\text{T}}$ is the number of training points and $b_{\text{T}}$ is the precision, in bits, of the training $y_i$ values). Once $T$ exceeds $C$, we see a dramatic increase in the training error, and the test and training error converge to similar values, evidencing that the learned model generalizes (at least over the trained domain, see Ref.~\cite{SM}).
	
	\textit{Conclusion:} We have introduced an information-theoretic measure constraining the learnable complexity of quantum learning machines, and have shown that it is bounded by the number of trainable parameters and their trainable precision. Our results apply to QNNs regardless of whether the trained parameters, learning method, or data are quantum or classical. We demonstrated these findings on the example of a Gaussian Boson Sampler-based feed-forward quantum reservoir computer and showed how the need to estimate expectation values (in this particular QNN) results in an exponential decrease in the capacity. While advantages in other metrics (e.g., training speed or performance on specific tasks \cite{quantumdomain}) are still possible, without quantum training QNNs will have limited advantages over classical NNs in terms of capacity. With quantum training however, QNNs that provide a capacity advantage over classical NNs seem to be possible. Ultimately, an exponential capacity advantage -- on both quantum and classical tasks -- may be achievable. 
	
	{\it Note:} During preparation of this manuscript, we noticed the posting of two related preprints \cite{Banchi2019,Dolzhkov2019}. {\it Acknowledgments:}
	We thank Tatsuhiro Onodera, Yihui Quek and Dmitri Pavlichin for helpful discussions. This work was partially supported by the Impulsing Paradigm Change through Disruptive Technologies (ImPACT) Program of
	the Council of Science, Technology and Innovation (Cabinet Office, Government of Japan). LGW acknowledges additional support from Cornell Neurotech.
	
	%\bibliography{QRC} 

\begin{thebibliography}{70}
\bibitem [{quantumdomainM()}]{quantumdomain}%
		\BibitemOpen
		\bibfield  {booktitle}  {\bibinfo {booktitle} {For example, even capacity-limited QNNs offer a natural advantage over classical NNs for operations on quantum data, since using a QNN avoids needing to first perform tomography to determine the classical representation of the input quantum state} }\BibitemShut {NoStop}%
\end{thebibliography}
	%\bibliographystyle{apsrev4-1}

	
	\onecolumngrid
	\appendix	
	
	\renewcommand\thefigure{\thesection \arabic{figure}}  
	
	\section{The Capacity of Quantum Neural Networks: Supplemental Material}
	
	\setcounter{figure}{0}    
	\setcounter{equation}{0}
	\subsection{Detailed description of the GBS-fQRC}
	Our GBS-fQRC reservoir circuit consists of $M$ input squeezers, followed by a random $M$-mode interferometer and finally, measurement in the Fock basis. Quantum data is input as states directly into the circuit. We input classical data, $x_i$, to the device by encoding it in the squeezing parameter of the squeezed-vacuum states (i.e., the $i$th squeezer is realized with an operator $\hat{S}_i \propto \exp[x_i(\hat{a}_i^2-(\hat{a}_i^{\dagger})^2)]$). The choice of a nonlinear embedding, rather than state-amplitude encoding, is an important part of transforming the input classical data into the reservoir circuit's high-dimensional Hilbert space. We consider a quantum-classical hybrid approach in which training and the final layer of the QNN is implemented with a classical computer. The output of the GBS-fQRC reservoir circuit is chosen to be a vector of photon-detection expectation values, including coincidences. There are $2^M-1$ such expectation values. For the 5-mode device we consider throughout, the reservoir output vector is $\vec{R}=[\langle \hat{n}_1 \rangle$,$\langle \hat{n}_2 \rangle$,...,$\langle \hat{n}_1\hat{n}_2 \rangle$,...,$\langle \hat{n}_1\hat{n}_2\hat{n}_3\hat{n}_4\hat{n}_5 \rangle]$. To vary $N_{\text{w}}$, we truncate $\vec{R}$ to the first $N_{\text{w}} \le 2^5-1$ elements. For execution of the trained QNN, this vector is multiplied by a trained output-layer matrix $W_{\text{out}}$, which contains all the trainable parameters of the fQRC, to produce the computation result. Training the fQRC requires collecting the output vectors $\vec{R}$ across a training set, consisting of input-output pairs $\vec{x}_{\text{T}}$ and $\vec{y}_{\text{T}}$. The reservoir output vectors are assembled into a matrix $R_{\text{out}}$, and the optimum $W_{\text{out}}$ is learned simply by $W_{\text{out}}=R_{\text{out}}^+\vec{y}_{\text{T}}$, where $R_{\text{out}}^+$ is the Moore-Penrose pseudo-inverse of $R_{\text{out}}$.
	
	In our simulations of GBS-fQRC presented here, we considered Fock states up to $\ket{3}$. To ensure that states remained consistently within this cutoff, we did not include any displacement of the input states, and allowed for a maximum squeezing of 2.6 dB for any input. The interferometers were all set to implement Haar-random unitaries, where for each task including training and testing and varying of parameters, we considered a fixed interferometer. We did not observe significant changes in results when different random unitaries were chosen (from the set of Haar-random unitaries). 
	
	\subsection{Achieving the capacity upper bound with fQRC}
	One reason we have chosen fQRC as our example system for numerically demonstrating our results is that the simple training permits us to easily show that $C$ may reach the maximum value. We are interested in the case where the exact mapping of $N$ distinct inputs each to one of $N$ distinct output $m$-bit numbers can be learned. In this case, we require that $W_{\text{out}}=R_{\text{out}}^{-1}\vec{y}_{\text{T}}$ have an exact solution. If $N_{\text{w}}$ is the number of $m$-bit outputs of the reservoir, then $R_{\text{out}}$ is an $N_{\text{T}}$-by-$N_{\text{w}}$ matrix, where $N_{\text{T}}$ is the number of training input-output pairs. $\vec{y}_{\text{T}}$ is a $N_{\text{T}}$-element vector. If the maximum number of \textit{distinct} input-output pairs, $N$, is less than $N_{\text{T}}$, then all but $N$ of the $N_{\text{T}}$ outputs can be assigned zero weight (ignored), such that finding $R_{\text{out}}^{-1}$ reduces to finding the inverse of a square $N$-by-$N$ matrix. If, for $N\le N_{\text{w}}$, the rank of $R_{\text{out}}$ is at least $N$, then an exact solution for $W_{\text{out}}$ exists. Provided this can be true for $N=N_{\text{w}}$, and that $W_{\text{out}}$ consists of $N_{\text{w}}$ $m$-bit numbers (i.e., $W=N_{\text{w}}m$) then the fQRC's capacity can maximal, that is: $C=N_{\text{w}}m=W$.
	
	In other words, to achieve maximum capacity with respect to $W$ with fQRC, the requirements are  that the parameters can be trained to at least $m$-bit precision, and that the independent features that make up the reservoir-output vector $\vec{R}$ (here different expectation values) correspond to linearly independent transformations of the input data over the training data set. For the choice of $\vec{R}$ considered in the main article and the specific GBS device details we chose, reaching high precision requires impractically many samples but is possible in principle. Provided all the chosen expectation values have finite values, they correspond to linearly independent transformations of the input (since any of the chosen operators cannot be written as a linear combination of the others).

	\subsection{Estimation of capacity}
	To estimate the capacity of neural networks we employed two approaches. In both approaches a training data set is constructed as follows. $N$ distinct random points distributed uniformly over the input space of the neural network were used as inputs, $\vec{x}_{\text{T}}$. The intended output for each input was then assigned to be one of $N$ random, uniformly-distributed $m$-bit numbers. In practice, here we exclusively use for our $m$-bit numbers Python double-precision numbers, since these are the values consistent with squeezing parameters and expectation values in our quantum simulations. 
	
	To produce the results here, we computed the mean training error across many sets of outputs. To determine with certainty a QNN's capacity $C$, one needs to ensure that all possible $N$, $m$-bit labellings can be learned. This is impractical for even small numbers: for our simple 5-mode GBS-fQRC the number of labellings is $\approx 2^{(2^5-1)\times 45}$, where $2^5-1$ is the number of distinct expectation values (the number of trainable parameters if the GBS-fQRC is used to produce scalar outputs as it is here), and $45$ bits is the effective precision of these trainable parameters in our simulations (see later in this section for more details). Thus, we instead choose to merely estimate $C$ by considering a small number, typically 10-100, of randomly-chosen labellings, as little difference was observed with larger sample sizes. We also confirmed that similar results could be obtained for unusual labellings, such as if the intended output $N$ $m$-bit number labels were all chosen to be identical, or when all but one were identical.
	
	As a baseline test of this capacity-estimation procedure, we first applied it to classical extreme learning machines (ELMs), which are single-hidden-layer neural networks with randomly-initialized parameters, trained with linear regression on the linear output layer. The classical ELMs considered here (and in the main manuscript) have the following design. The networks had 5 input nodes, $N_{\text{w}}$ hidden nodes, and 1 output node. Input data was normalized to $\pm 1$ and hyperbolic tangent activation was used with zero bias. An ELM has a linear input layer, in which inputs are distributed to the hidden layer by multiplying the input vector by a random matrix, $W_{\text{in}}$. The matrix elements of the input layer to the ELMs were chosen from a uniform random distribution, and normalized to 1 by default. For evaluating performance of the ELM on various functions, we additionally varied the normalization by multiplying $W_{\text{in}}$ by adjusting a scalar multiplier $\rho$ (see later section of this document). 
	
	To assess the effect of noise in the read-out of the hidden layer (similar to the estimation-associated noise in the GBS-fQRC or other QNN), we added mean-zero noise from a uniform distribution with varying peak amplitude each time the activation of the hidden layer was calculated (i.e., the noise added to the output of the hidden layer every time it was executed, both in training and testing, was different). We also used this test to determine the numerical-noise-related precision of our calculations.
	
	To estimate $C$ we first calculated the absolute normalized error after training, $\epsilon (N) = \text{mean}[\frac{|y_i-y_{ti}|}{|y_{ti}|}]$, where $y_i$ is the NN output after training for $N$ inputs/labels, $y_{ti}$ is the intended training output (the $N$ labels), and the mean is taken over both the individual training set and all the random labellings considered. From this, we then determined $C$ as
	
	\begin{equation} \label{1}
	C=\text{max}[N\log_2(1/\epsilon(N))],
	\end{equation} 
	
	\noindent where the maximum is taken over different values of $N$. When this is done, we obtain the results shown in Supplementary Figure 1.
	
	\begin{figure}
		\centering
		\includegraphics[width=6 in]{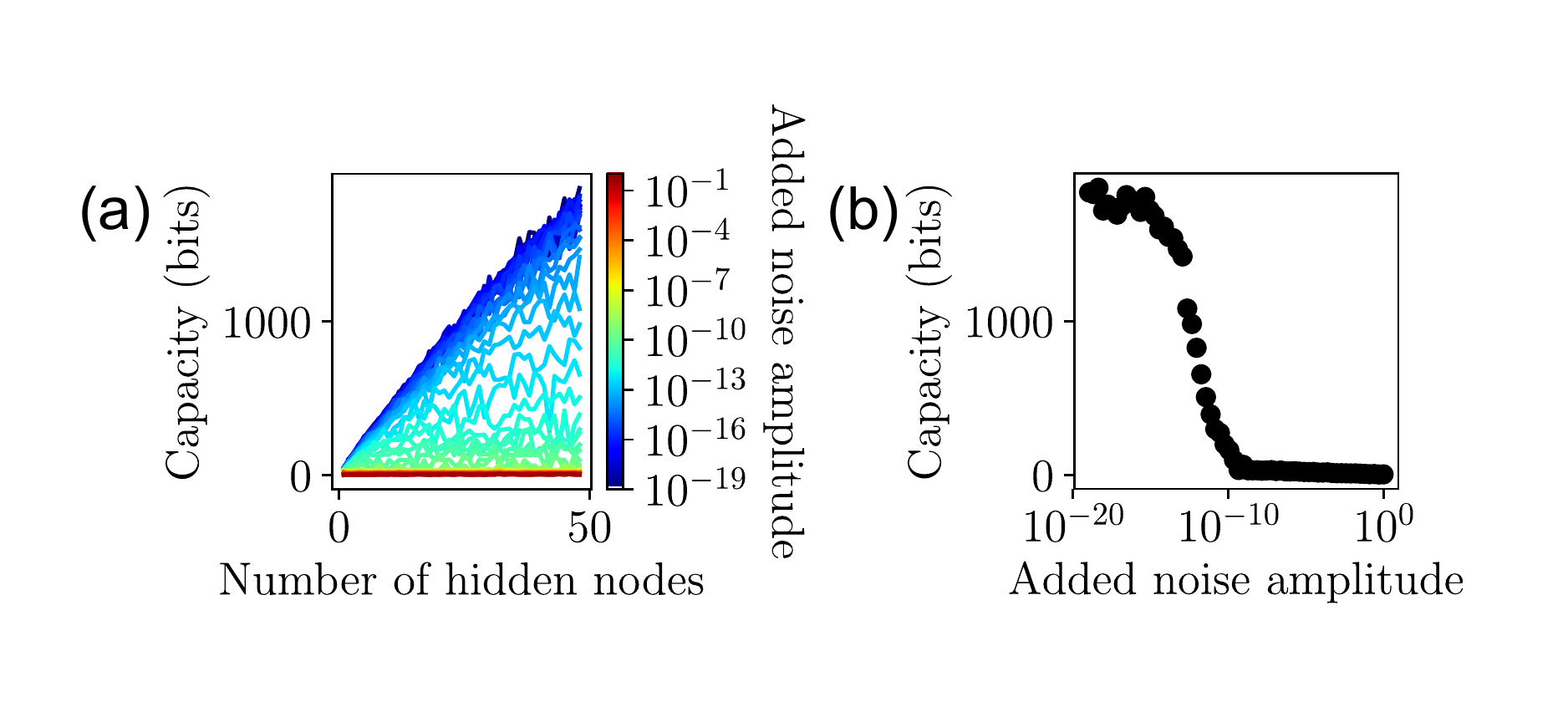}
		\caption{Direct estimation of the capacity. (a) Estimated capacity for a classical single hidden layer neural network (extreme learning machine). Different curves correspond to different additive noise levels. (b) The estimated capacity (for 48 hidden nodes/trainable parameters) versus the additive noise amplitude. }
	\end{figure}
	
	Supplementary Figure 1a shows that, while there is typically a linear or sub-linear growth of $C$ with more trainable parameters, even with zero artificial additive noise none of the tested neural networks reach the maximum possible capacity. Supplementary Figure 1b clarifies this finding as resulting from the limited numerical precision of our simulations. We see that, for artificial additive noise up to about $10^{-14}$, no change in the estimated capacity occurs. This shows that the calculations (including training, and thus $W$) are limited by a numerical-precision noise to some finite precision we denote $\epsilon_{\text{p}}$ (measured in units of bits); the source of this noise is in general a combination of the finite-precision representation of numbers presumed real by some of the calculations, and the propagated error from those calculations. Based on the curves in Supplementary Figures 1b and 2b, we estimate $\epsilon_{\text{p}}$ to be $\log_2(1/10^{-14}) \approx 46.5$ bits. Once the additive noise is larger than this intrinsic noise, we see the expected logarithmic decay of the capacity (approximately linear on the log-x axis of Supplementary Figure 1b), until at around $10^{-8}$, zero capacity is reached. This point corresponds to the point of zero signal-to-noise ratio in training of the parameters.
	
	\begin{figure}
		\centering
		\includegraphics[width=6 in]{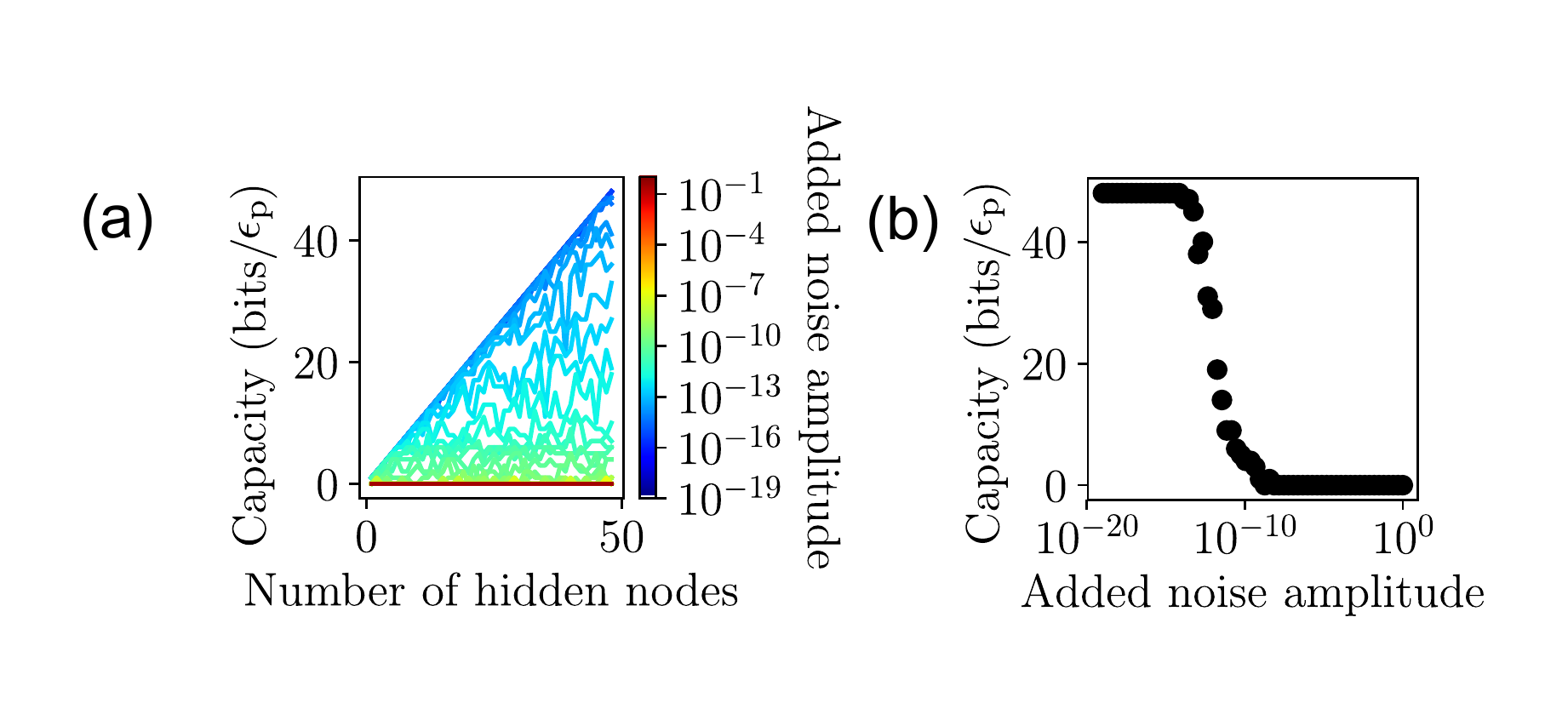}
		\caption{Numerical-noise-normalized capacity estimate. (a) Estimated capacity for a classical single hidden layer neural network (extreme learning machine) in units of simulation precision $\epsilon_{\text{p}}$. Different curves correspond to different additive noise levels. (b) The estimated capacity (for 48 hidden nodes/trainable parameters) versus the additive noise amplitude. }
	\end{figure}
	
	The previous results show that our estimate of $C$, via the direct Eqn. \ref{1}, has the feature of being sensitive to the precision of the calculations. This is correct: $C$ itself depends on the precision of the calculations that define the learning machine's training and execution. However, here we are interested in evaluating $C$ due to the physics of a simulated machine, not due the finite precision of the simulation. In the interest of concentrating on the physical aspects of the network that affect $C$, we chose to also measure $C$ relative to the numerical precision $\epsilon_{\text{p}}$. To do this, we simply take $C=\text{max}[N]\times \epsilon_{\text{p}}$ for networks where the mean error is smaller than $10^{-10}$ (a normalized mean squared error of $10^{-20}$). Supplementary Figure 2 shows the result of applying this policy for increasing added noise amplitude. 
	
	In the main article the latter policy is used to estimate the capacity of the GBS-fQRC when the exact expectation values are used in $\vec{R}$, in order to present the physics-limited capacity (within the assumption of arbitrary-precision in measuring the expectation values) rather than one limited by the precision of our simulations. However, Eqn. \ref{1} and the associated procedure are the most general approach for estimating $C$, and are used otherwise. 
	
	\subsection{Other examples of classical and quantum tasks}
	
	\begin{figure*}
		\centering
		\includegraphics[width=\textwidth]{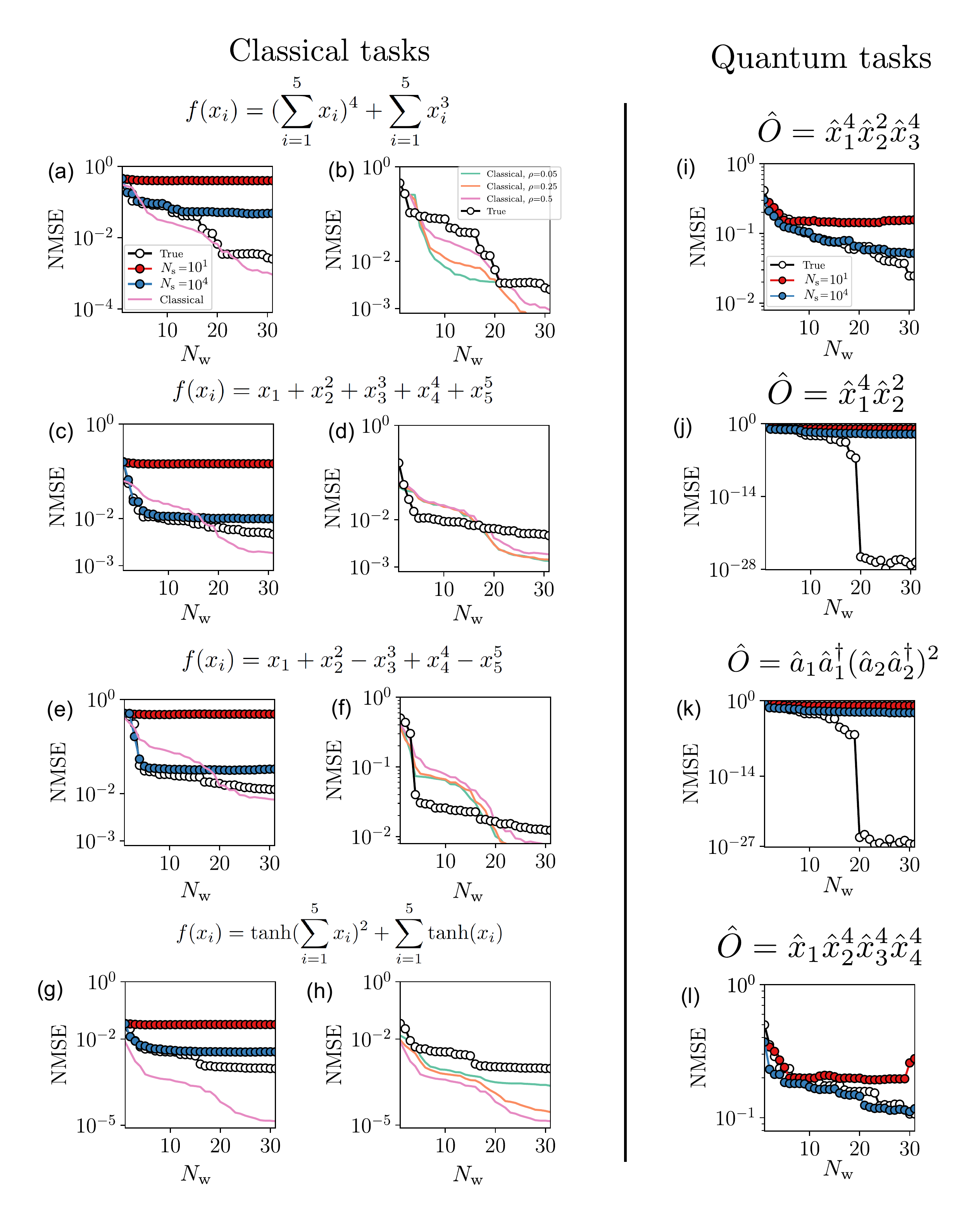}
		\caption{Additional examples of classical and quantum function approximation. For details see text. }
	\end{figure*}
	
	As mentioned in the main article, we see similar trends for a variety of quantum and classical continuous-function-approximation tasks. Supplementary Figure 3 shows a collection of these tasks. In each case, we consider a 5-mode GBS with 5 inputs. For training data, the intended function (operator-expectation-value, in the quantum case) was calculated for 500 random sets of $x_i$ (classical) or 500 random states (quantum).  In the quantum case, input states were of the form $\ket{\psi_1} \otimes \ket{\psi_2} \otimes \cdots \otimes \ket{0}$ where each $\ket{\psi_i}$ is randomly selected to be a Fock, cat, coherent, or thermal state, with a randomly-selected mean photon number between 0 and 1 (or equal to 1 or 2 for the Fock state). The number of non-vacuum states for each task was determined by the number of different modes in the operator (e.g., in Supplementary Fig 3i, 3 non-vacuum states were combined with 2 vacuum states to form the input state $\ket{x}$). For both types of tasks, the performance was measured on an independent verification (i.e., test) set of 500 independently drawn random inputs, where the normalized mean squared error (NMSE) was calculated with respect to the correct function (operator expectation) value. As is typical for machine learning, the test data is drawn from the same range of values as the training data (i.e., the functions are learned over a finite domain, and generalization ability is assessed only over the same domain). For the classical tasks, we compare the fQRC with a classical extreme learning machine. This classical ELM has a hyperparameter in its $\tanh$ activation function, $\rho$, that scales the inputs, i.e., $\vec{R}_{\text{classical}}=\tanh(\rho W_{\text{in}}\vec{x})$, where $W_{\text{in}}$ is the random linear input-layer matrix, and $\vec{x}$ is the vector of input values. For completeness, the second column of Supplementary Figure 3 shows the classical NMSE for different values of this hyperparameter. Although the performance in the classical case can vary strongly with this parameter, we generally find that the performance of the classical and quantum NNs with true expectation values are similar when they have a similar number of parameters. Quantum NNs with estimated expectation values show worse performance than classical ELMs with similarly-many parameters.
	
	In general, for most tasks we consider, the same trends reported in the main article are observed: better performance is obtained with more trainable parameters and with more samples used to estimate expectation values in $\vec{R}$. For completeness however, we note that these are merely overall trends. We sometimes notice that particular features (expectation values in $\vec{R}$) lead to dramatically improved performance (e.g., Supplementary Figure 3j). This is to be expected when certain features are strongly correlated with the intended function. In addition, when the complexity of tasks is too high, the learned model is sufficiently far from the intended function that we often see unexpected trends, such as monotonic increases in performance with larger numbers of parameters $N_\text{w}$ or samples $N_\text{s}$ (e.g., Supplementary Figure 3l).
	
	\subsection{Definition of NMSE}
	The definition of the normalized mean squared error we use is:
	\begin{equation}
	\text{NMSE} := \frac{1}{N}\sum_{i=1}^{N}\frac{(y_i-y_{\text{v},i})^2}{(y_{\text{v},i})^2},
	\end{equation}
	
	\noindent where $N$ is the number of samples in the validation test set (or the training set, if the NMSE refers to training error), and $y_{\text{v},i}$ are the correct output values in the validation test (training) set.
	
	\subsection{More on generalization and capacity}
	
	In the main article, we consider the generalization performance as a function of training data within a fixed range of input values. To better visualize the trade-offs between capacity and generalization, here we consider two 1-dimensional functions, and examine the effect of capacity on generalization \textit{outside} the range of inputs used for training (Supplementary Figure 4). 
	
	\begin{figure*}
		\centering
		\includegraphics[width=\textwidth]{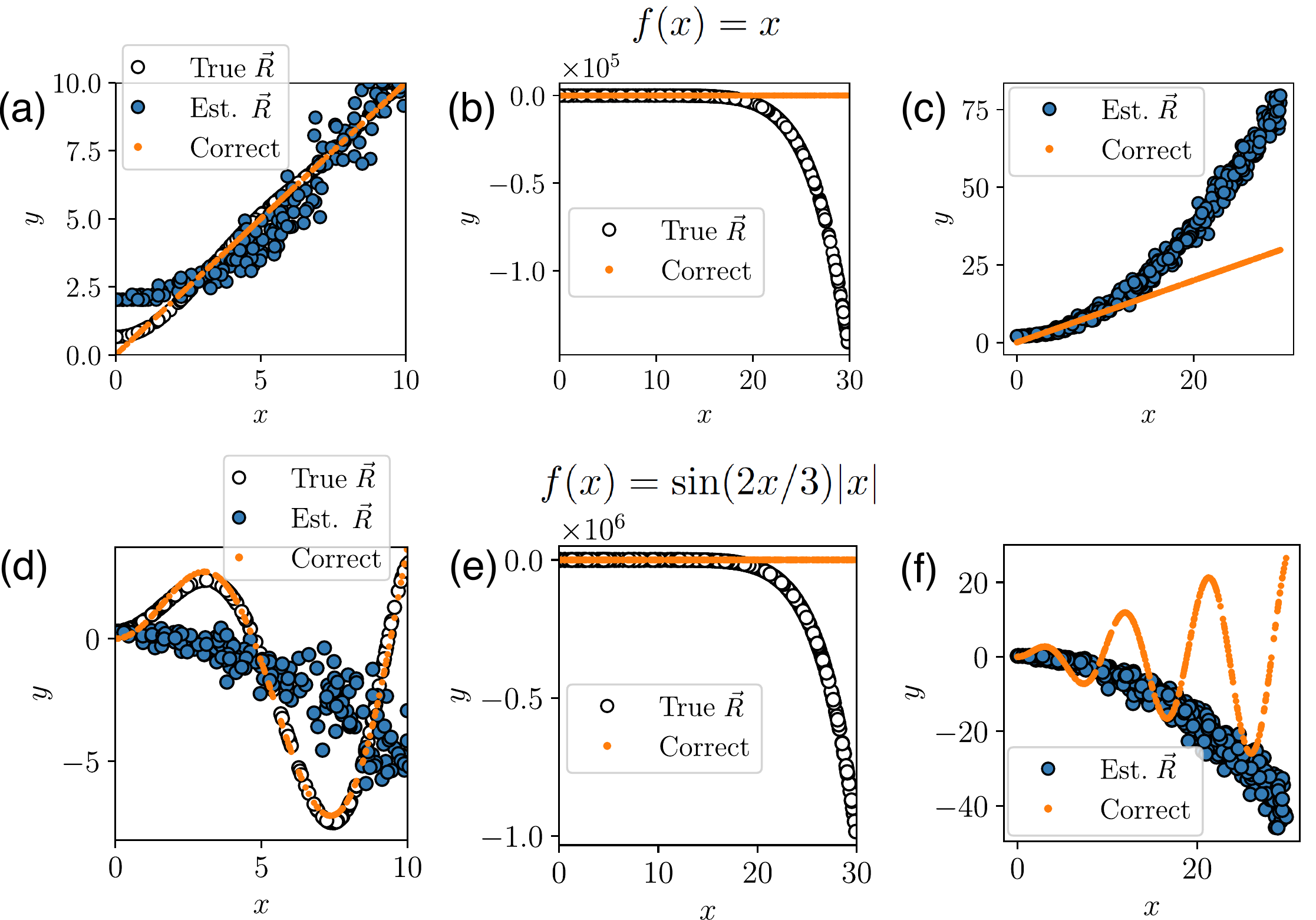}
		\caption{Test performance of a high-capacity QNN (GBS-fQRC using exact simulated expectation values) and a low-capacity QNN (GBS-fQRC using estimated expectation values) on 1-dimensional function approximation. For both the simple linear function and the more complex sinusoidal function, significantly better performance is achieved by the high-capacity model within the training data range, $x \in [0,10]$. For the complex function, the difference in performance is more noticeable. However, for testing with $x$ outside the range $[0,10]$, the high-capacity model's predictions diverge much more quickly from the correct answers than do the low-capacity model's ones. The experiment was set up such that, in both cases, the training data significantly exceeds the capacity. The panels from left (a,d) to right (c,f) show the test performance over the range of the training set (a,d), the test error over the extended test range for the case where exact expectation values are used (b,e), and when the estimated expectation values are used (c,f).}
	\end{figure*}
	
\end{document}